\documentclass[twocolumn,showpacs,preprintnumbers,amsmath,amssymb,superscriptaddress]{revtex4}

\usepackage{epsf}
\usepackage{graphicx}
\usepackage{sidecap}

\usepackage{color}

\def \beq {\begin{equation}}
\def \eeq {\end{equation}}
\begin{document}

\title{{Non-Kondo-like Electronic Structure in the Correlated Rare-Earth Hexaboride YbB$_6$}}



\author{Madhab~Neupane}
\affiliation {Joseph Henry Laboratory and Department of Physics,
Princeton University, Princeton, New Jersey 08544, USA}

\author{Su-Yang~Xu}
\affiliation {Joseph Henry Laboratory and Department of Physics, Princeton University, Princeton, New Jersey 08544, USA}

\author{Nasser~Alidoust}\affiliation {Joseph Henry Laboratory and Department of Physics, Princeton University, Princeton, New Jersey 08544, USA}

\author{Guang~Bian}\affiliation {Joseph Henry Laboratory and Department of Physics, Princeton University, Princeton, New Jersey 08544, USA}

\author{D.J. Kim}
\affiliation {Department of Physics and Astronomy, University of California at Irvine, Irvine, CA 92697, USA}

\author{Chang~Liu}
\affiliation {Joseph Henry Laboratory and Department of Physics,
Princeton University, Princeton, New Jersey 08544, USA}

\author{I.~Belopolski}
\affiliation {Joseph Henry Laboratory and Department of Physics, Princeton University, Princeton, New Jersey 08544, USA}

\author{T.-R. Chang}
\affiliation{Department of Physics, National Tsing Hua University, Hsinchu 30013, Taiwan}

\author{H.-T. Jeng}
\affiliation{Department of Physics, National Tsing Hua University, Hsinchu 30013, Taiwan}
\affiliation{Institute of Physics, Academia Sinica, Taipei 11529, Taiwan}

\author{T.~Durakiewicz}
\affiliation {Condensed Matter and Magnet Science Group, Los Alamos National Laboratory, Los Alamos, NM 87545, USA}

\author{H.~Lin}
\affiliation {Graphene Research Centre, Department of Physics, National University of Singapore, Singapore 117542, Singapore}


\author{A.~Bansil}
\affiliation {Department of Physics, Northeastern University,
Boston, Massachusetts 02115, USA}

\author{Z. Fisk}
\affiliation {Department of Physics and Astronomy, University of California at Irvine, Irvine, CA 92697, USA}

\author{M.~Z.~Hasan}
\affiliation {Joseph Henry Laboratory and Department of Physics,
Princeton University, Princeton, New Jersey 08544, USA}
\affiliation {Princeton Center for Complex Materials, Princeton University, Princeton, New Jersey 08544, USA}

\date{today}
\begin{abstract}

We present angle-resolved photoemission studies on the rare-earth hexaboride YbB$_6$, which has recently been predicted to be a topological Kondo insulator. Our data do not agree with the prediction and instead show that YbB$_6$ exhibits a novel topological insulator state in the absence of a Kondo mechanism. We find that the Fermi level electronic structure of YbB$_6$ has three 2D Dirac cone like surface states enclosing the Kramers' points, while the $f$-orbital which would be relevant for the Kondo mechanism is $\sim1$ eV below the Fermi level. Our first-principles calculation shows that the topological state which we observe in YbB$_6$ is due to an inversion between Yb $d$ and B $p$ bands. These experimental and theoretical results provide a new approach for realizing novel correlated topological insulator states in rare-earth materials.

\end{abstract}
\date{\today}
\maketitle

Rare-earth materials are interesting because the strong electronic correlation in their $f-$electrons leads to novel ground states such as a Kondo insulating state, valence fluctuation state, and heavy fermion superconductivity state \cite{Fisk, Coleman, Riseborough, Antonov}. Among these correlation-driven ground states, Kondo insulators are characterized by a particularly narrow band gap (on the order of 10-40 meV) at low temperatures with the chemical potential in the gap. Unlike typical band insulators the energy gap that opens up at low temperatures in a Kondo insulator is in fact due to the Kondo hybridization of localized $f-$electrons with conduction electrons \cite{Fisk, Riseborough, Coleman,Antonov}. With the advent of topological insulators \cite{Hasan, Hsieh, SCZhang, FuKaneMele, Xia, Hasan2, Neupane_1}, recently the rare-earth Kondo insulator, SmB$_6$, has attracted much interest due to the theoretical prediction that it exhibits a topological Kondo insulator (TKI) phase \cite{Dzero,Takimoto, Dai}. In a TKI, the Kondo hybridization between the $d$ and $f$ bands further leads to an inversion of the band parity, and therefore realizes a topologically nontrivial insulator phase. Following the theoretical predictions, photoemission and transport experiments have identified the existence of an odd number of in-gap surface states and a two-dimensional conductance channel at low temperatures in SmB$_6$ \cite{Neupane, DLFeng, Fisk_discovery, Hall, tunneling, Fisk_1, Nan}, consistent with the theoretically predicted TKI phase. However, the surface states in TKI phase of SmB$_6$ only exist at very low temperatures \cite{Neupane} and their Fermi velocity is expected to be low due to a strong $f$-orbital contribution \cite{Dai, Neupane}, limiting its future applications in devices.

\begin{figure}
\centering
\includegraphics[width=9cm]{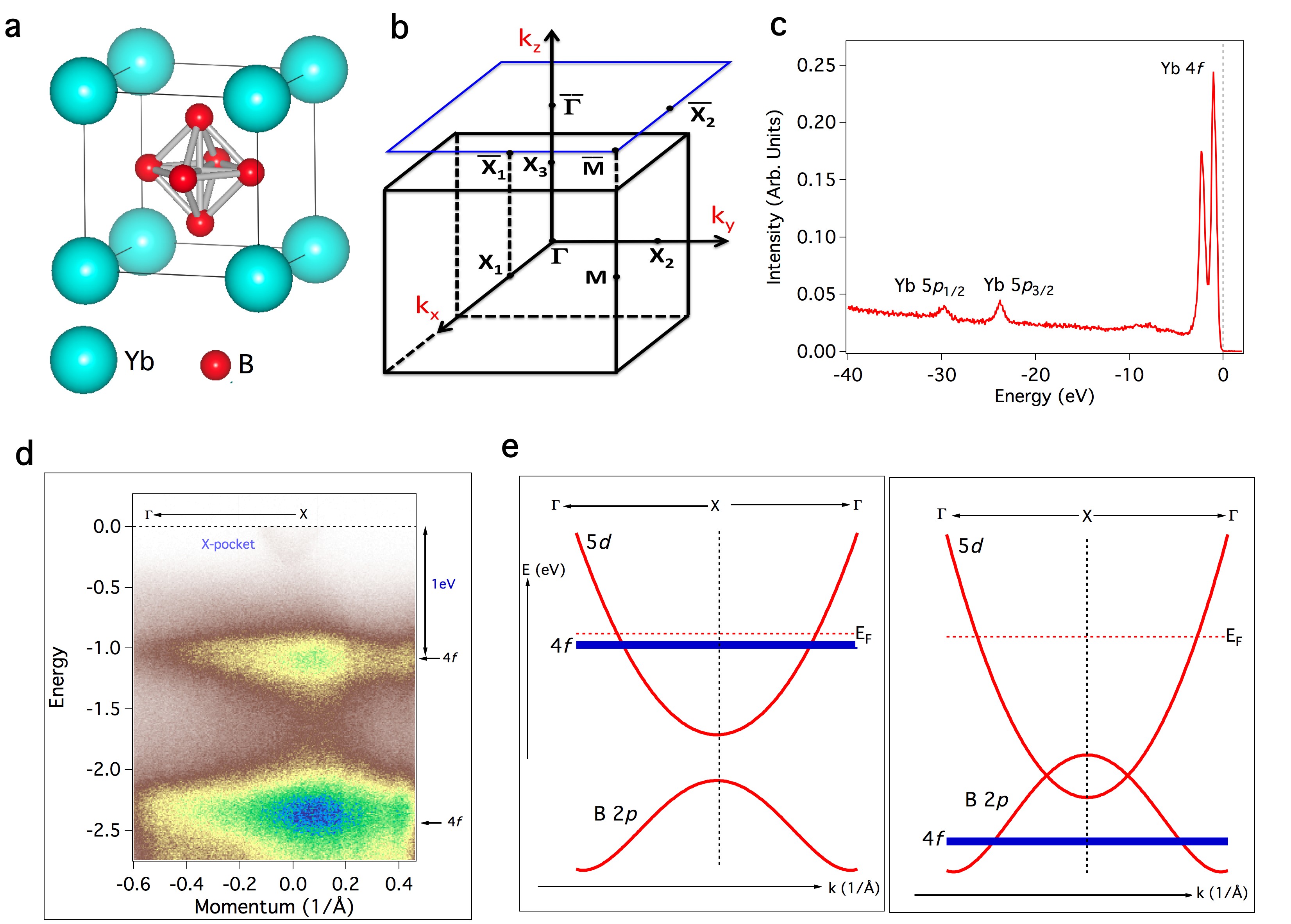}
\caption{{Brillouin zone symmetry and band structure of YbB$_6$}. (a) Crystal structure of YbB$_6$. Yb ions and B$_6$ octahedron are located at the corners and the center of the cubic lattice structure. (b) The bulk and surface Brillouin zones of YbB$_6$. High-symmetry points are marked. (c) Core-level spectroscopic measurement of YbB$_6$. Various energy levels are marked on the curve. (d) ARPES measured dispersion maps along the ${{\bar\Gamma}-\bar{X}-{\bar\Gamma}}$ momentum-space cut-directions. Dispersive cone like pocket and non-dispersive flat Yb 4$f$ bands are observed. This spectrum is measured with photon energy of 32 eV at temperature of 15 K in SRC PGM beamline. (e) Different types of possible electronic ground state in rare-earth hexaborides.}
\end{figure}

\begin{figure*}
\centering
\includegraphics[width=17.0cm]{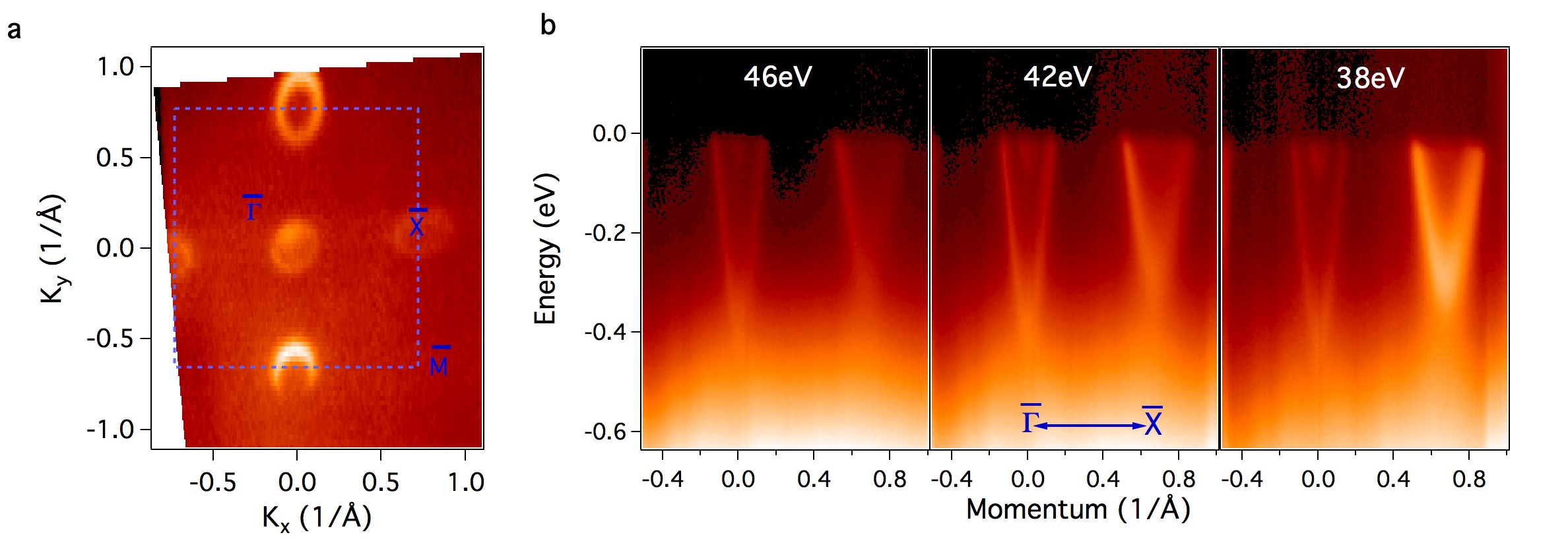}
\caption{{Fermi surface and dispersion map.} 
(a) ARPES measured Fermi surface of YbB$_6$. Circular shaped pockets are observed at ${\bar\Gamma}$ and $\bar{X}$ points. 
It is measured with photon energy of 50 eV at temperature of 15 K. (b) ARPES dispersion maps measured with different photon energy. The measured photon energies are noted on the plots. These data were collected at ALS BL10.}
\end{figure*}

\begin{figure}
\centering
\includegraphics[width=9cm]{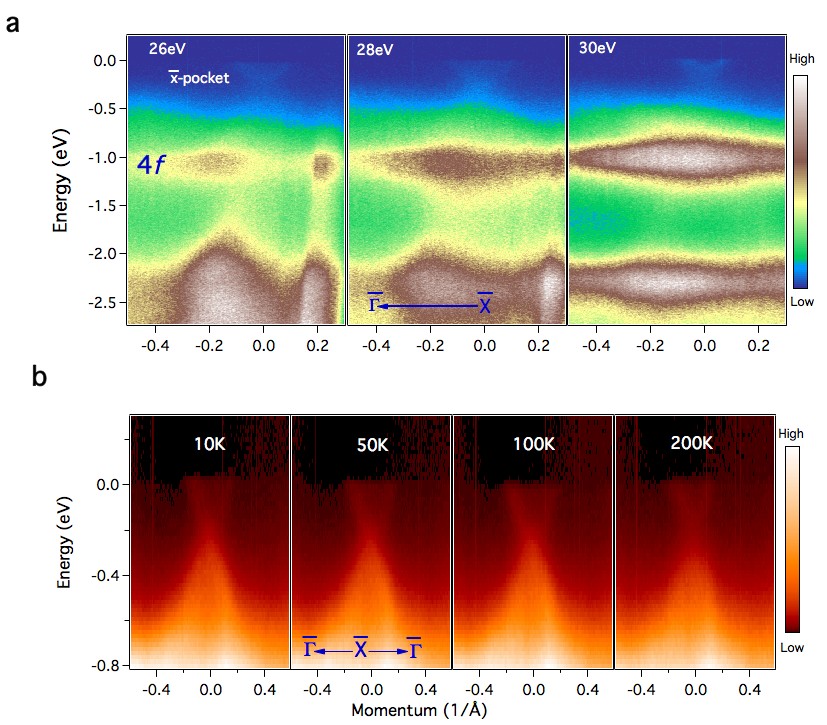}
\caption{{Photon energy and temperature dependent dispersion maps } (a) Photon energy dependent dispersion map with wider binding energy. The measured photon energy are noted on the plots. The 4$f$ flat bands are about 1 eV below the Fermi level.  These spectra are measured along the $\bar{\Gamma}- {\bar{X}}$ momentum space cut direction with temperature of 15K. These data were collected at SRC PGM beamline. (b) Temperature dependent ARPES spectra. The measured temperatures are noted on the plots. These spectra are measured along the $\bar{\Gamma}- {\bar{X}}-\bar{\Gamma}$ momentum space cut at ALS BL10.0.1.}
\end{figure}

In order to search for other novel correlated topological phases even without a Kondo mechanism, it is quite suggestive to systematically study other rare-earth materials that are closely-related to SmB$_6$. SmB$_6$ belongs to a class of materials called the rare-earth hexaborides, RB$_6$ (R $=$ rare-earth metal). In general, rare-earth hexaboride compounds can have three types of electronic bands in the vicinity of the Fermi level, namely the R $5d$ orbital band, the R $4f$ orbital band, and the boron $2p$ band. The low energy physics for a rare-earth hexaboride is collectively determined by the relative energies between these bands and the Fermi level, which depends on a delicate interplay among the key physical parameters including the valence of the rare-earth element, the lattice constant, the spin-orbit coupling (SOC), etc. As a result, the rare-earth hexaborides can realize a rich variety of distinct electronic ground states such as the proposed TKI state in SmB$_6$, ferromagnetic order in EuB$_6$ and superconductivity in LaB$_6$ ($T_c\sim0.5$ K) \cite{Dai_1, Antonov, resis}.


Despite these interesting aspects, apart from SmB$_6$, experimental studies on the electronic groundstates of RB$_6$ compounds are almost entirely lacking. In contrast to the Kondo insulator state in SmB$_6$, YbB$_6$ is known to be a doped semiconductor from transport experiments \cite{resis}. This difference is particularly interesting since SmB$_6$ and YbB$_6$ share many important properties such as the same crystal structure and the same sets of low energy bands (R $5d$, R $4f$, and B $2p$). Moreover, it is interesting to investigate whether the semiconducting state in YbB$_6$ can also host topological order and how that order is different from that in the TKI phase in SmB$_6$. Here we report systematic studies on the electronic groundstate of YbB$_6$. We show that unlike in SmB$_6$, the lowest $f$-orbital in YbB$_6$ is about 1 eV away from the Fermi level suggesting the near absence of Kondo insulating behavior in contrast to the well known mixed-valence system SmB$_6$. Our spectroscopic data show that the Fermi level of YbB$_6$ features nearly linearly dispersive states without observable out-of-plane momentum $k_z$ dispersion. Guided by the systematic experimental data, we present new band calculations with an adjustable correlation parameter (Hubbard-$U$) which reveal that the observed topological state in YbB$_6$ is due to a band inversion between the $d$ and $p$ bands under a nonzero Coulomb interaction value.

Single crystal samples of YbB$_6$ were grown in the Fisk lab at the University of California (Irvine) by the Al-flux method, which is detailed elsewhere \cite{Fisk_discovery,resis}. Synchrotron-based angle-resolved photoemission spectroscopy (ARPES) measurements of the electronic structure were performed at the PGM beamline of the Synchrotron Radiation Center, Stoughton, WI and Beamline 10.0.1 of the Advanced Light Source, Berkeley, CA equipped with high-efficiency Scienta R4000 electron analyzers. The energy resolution was 10-30 meV and the angular resolution was better than 0.2$^{\circ}$ for all synchrotron measurements. The samples were cleaved along the (001) plane and were measured in ultrahigh vaccum better than 10$^{-10}$ Torr. The first-principles bulk band calculations were performed based on the generalized gradient approximation (GGA) \cite{GGA} using the projector augmented-wave method \cite{PAW, PAW_1} as implemented in the VASP package \cite{VASP, VASP_1}. The experimental crystallographic structure was used \cite{expt} for the calculations. The spin-orbit coupling was included self-consistently in the electronic structure calculations with a 12$\times$12$\times$12 Monkhorst-Pack $k-$mesh. The surface electronic structure computation was performed with a symmetric slab with a thickness of 24 unit cells.

YbB$_6$ shares the same CsCl type of crystal structure as SmB$_6$, as shown in Fig. 1a. The bulk Brillouin zone (BZ) is cubic, where the center of the BZ is the $\Gamma$ point and the center of each face is the $X$ point. It is known from theoretical calculations \cite{Antonov} that three different atomic orbitals contribute to the low-energy electronic structure, namely the the rare-earth $5d$ orbital band, the rare-earth $4f$ orbital band and the boron $2p$ band. Both the valence band maximum and the conduction band minimum are around the $X$-point \cite{Antonov}. A variety of electronic ground states can be realized due to the delicate interplay of the relative energy levels among these bands and the Fermi level. For example, the left panel of Fig. 1e shows a condition for the Kondo insulator state, as observed in SmB$_6$. The R $5d$ and R $4f$ bands cross each other near the Fermi level, whereas the B $2p$ band lies energetically below the $5d$ and $4f$ bands, and therefore is irrelevant to the low-energy physics. At low temperatures, the itinerant $5d$ electrons hybridize with the localized $4f$ electrons and open up a Kondo energy gap, which leads to the Kondo insulating state. On the other hand, the right panel of Fig. 1e presents a different scenario, where the R $4f$ band resides far away from the Fermi level. In this case, the R $5d$ and B $2p$ bands dominate the low-energy physics, and they can even show a finite band inversion depending on details of material parameters.  In order to reveal the electronic state of YbB$_6$, we systematically study its electronic structure at the (001) natural cleavage surface. Fig. 1c shows the momentum-integrated photoemission spectrum over a wide binding energy window. The Yb $4f$ and Yb $5p$ core levels are clearly observed. To more precisely determine the energy positions of the Yb $4f$ with respect to the Fermi level, we present a $k$-resolved dispersion map. As clearly seen in Fig. 1d, the lowest $4f$ flat band in YbB$_6$ is 1 eV below the Fermi energy (also see Supplementary Materials). This is in sharp contrast to the ARPES data on SmB$_6$, where the flat $4f$ band is found to be only $\leq15$ meV away from the Fermi level \cite{Neupane}. Therefore, our data reveal the physical origin for the absence of the Kondo insulating state in YbB$_6$, which also negates the recent theoretical work that predicts the existence of the topological Kondo insulator phase in YbB$_6$ \cite{Dai_1}. Apart from the intense $4f$ band, our data in Fig. 1d also reveal an electron-like pocket centered at the $\bar{X}$ point that crosses the Fermi level.

\begin{figure}
\centering
\includegraphics[width=8cm]{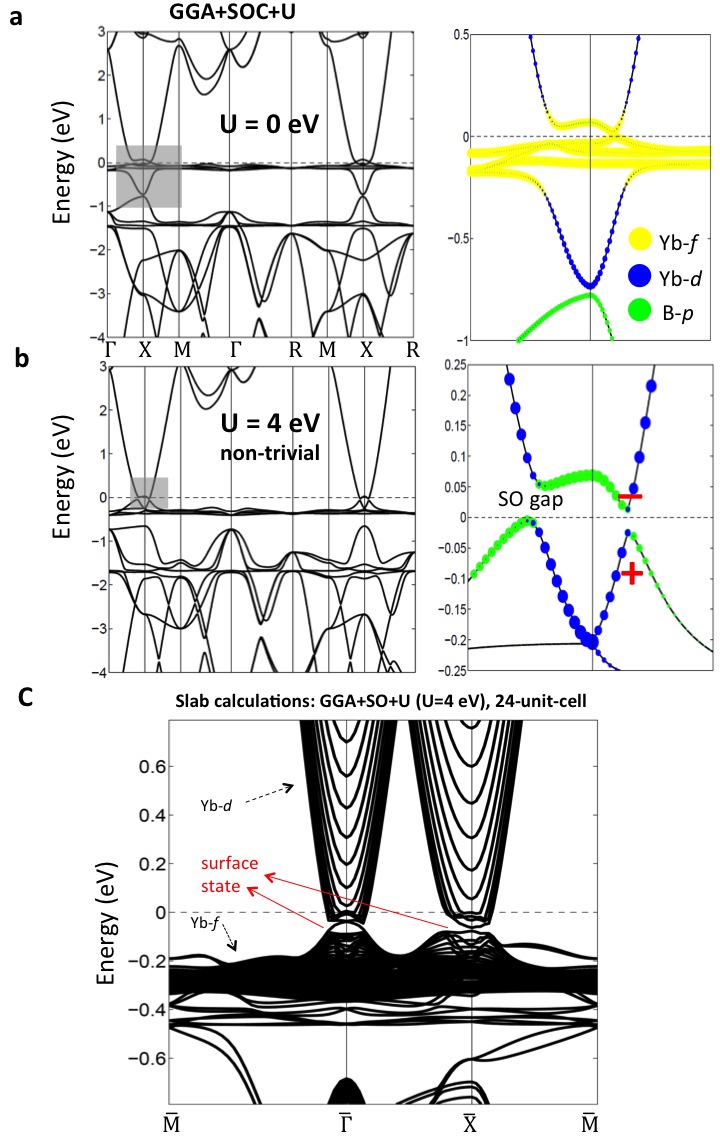}
\caption{{First-principle electronic structure of YbB$_6$.} (a) Theoretical bulk band structure along different high-symmetry points with on site Coulomb repulsion U = 0 eV. At U =0, the $f$ bands are closer to the Fermi level (see Fig a, right). (b) Same as (a) but with U = 4 eV, which shows the hybridization between $d$ and $p$ bands at $X$ point and consequently the gap is opened (see Fig b, right). The $+$ and $-$ signs show the parity eigen values of the bands at $X$ point, where a band inversion is found. (c) Slab calculations showing surface states in YbB$_6$.}
\end{figure}

In order to systematically resolve the Fermi level electronic structure, in Figs. 2 and 3 we present high-resolution ARPES measurements in the close vicinity of the Fermi level. It is important to note that at the studied (001) surface, the three $X$ points ($X_1$, $X_2$, and $X_3$) in the bulk BZ project onto the $\bar{\Gamma}$ point and the two $\bar{X}$ points on the (001) surface BZ, respectively. Since the valence band maximum and conduction band minimum are at the $X$-points, at the (001) surface one would expect low energy electronic states near the $\bar{\Gamma}$ point and $\bar{X}$ points.
The Fermi surface map of YbB$_6$ is presented in Fig. 2a. Our Fermi surface map reveals multiple pockets, which consist of an oval-shaped contour and a nearly circular-shaped contour enclosing each $\bar{X}$ and $\bar{\Gamma}$ points, respectively. No pocket is seen around the $\bar{M}$-point. YbB$_6$ shows metallic behavior in transport \cite{resis} which is consistent with our data.

We present ARPES energy and momentum dispersion maps, where low-energy states consistent with the observed Fermi surface topology are clearly identified. As shown in Fig. 2b, a ``V''-shaped linearly-dispersing band is observed at each $\bar{\Gamma}$ and $\bar{X}$ points. We further study the photon energy dependence of the observed ``V''-shaped bands. As shown in Figs. 2b and 3a, the ``V''-shaped bands are found to show no observable dispersion as the incident photon energy is varied, which suggests its two-dimensional nature. Therefore, our systematic ARPES data has identified three important properties in the YbB$_6$ electronic structure: (1) Odd number of Fermi surface pockets are observed to enclose the Kramers' point; (2) The bands at the Fermi level are found to exhibit nearly linearly (Dirac like) in-plane dispersion; (3) No observable out-of-plane ($k_z$) dispersion is observed for these Dirac like (``V''-shaped) bands. All these properties that we observed are consistent with a topological insulator state in YbB$_6$. We also find that the Dirac surface states are robust to the rise of temperatures, see Fig. 3b. This contrasts sharply with the in-gap state observed in SmB$_6$, which are observed to disappear above $\sim 30K$ \cite{Neupane}. Our systematic data are consistent with two other concurrent ARPES studies on YbB$_6$ \cite{YbB6_DLFeng, YbB6_Shi}.



In order to better understand the electronic structure that we observe in ARPES, we perform first-principles calculations on the bulk band structure of YbB$_6$ using the GGA + SOC + $U$ method. Our ARPES data shown here negate the predicted topological Kondo insulator state in a previous calculation result \cite{Dai_1}. Thus there is no theoretical calculation that can reproduce the novel topological insulator state in YbB$_6$ revealed by our ARPES data (also the two occurrent ARPES studies \cite{YbB6_DLFeng, YbB6_Shi}). Here, we present new band calculations with an adjustable correlation parameter (Hubbard-$U$). Figs. 4a and b show the calculated YbB$_6$ bulk band structure obtained by using various on-site $U$ values for the Yb $4f$ orbitals. For $U=0$, our calculation shows that the Yb $4f$ orbital is located very close to the Fermi level and crosses with the Yb $5d$ band. This is inconsistent with our data, which clearly shows that the lowest $4f$ band is 1 eV away from the Fermi level and does not cross the Yb $5d$ conduction band. Thus we study the calculated band structure with nonzero $U$ values. Fig. 4b shows the calculation for $U=4$ eV. Interestingly, it can bee seen that the Yb $4f$ band is moved toward higher energies away from the Fermi level as one increases $U$. For $U=4$ eV, the $4f$ band moves to a higher energy and does not cross the $5d$ band, which agrees with our data. Furthermore, the B $2p$ band is ``pushed'' toward the Fermi level and form a hole-like valence band, also consistent with our data. A finite band inversion between the Yb $5d$ and B $2p$ bands is obtained from our calculation for $U=4$ eV. Spin-orbit coupling opens a finite (inverted) energy gap, leading to an inversion of the parity eigenvalues at the $X$ point between the $d$ and $p$ bands (also see Supplementary Material). We conduct surface (slab) electronic structure calculation for the Coulomb interaction value $U=4$ eV, where a $d$-$p$ band inversion is seen. As shown in Fig. 4c, our slab calculation clearly shows topological Dirac surface states at both the $\bar{\Gamma}$ and $\bar{X}$ points. Furthermore the $4f$ bands are far away from the Fermi level and hence the Kondo mechanism is irrelevant. These systematic calculations, for the first time, reveal that the experimentally observed topological state in YbB$_6$ is due to a bulk band inversion between the $d$ and $p$ bands under a nonzero Coulomb interaction value.

In conclusion, by using high-resolution ARPES we show that, unlike SmB$_6$, the lowest $f$-orbital in YbB$_6$ is about 1 eV away from the Fermi level, which reveals that it is not a TKI. Our data show that the Fermi level of YbB$_6$ features linearly dispersive bands without observable out-of-plane momentum $k_z$ dispersion. These bands are found to form odd Fermi surface pockets enclosing the Kramers' points per BZ. We provide a theoretical calculation that reveals a novel topological insulator state due to band inversions between Yb $5d$ - B $2p$ orbitals, which is consistent with our data. Our results provides a new approach for realizing a topological insulator in rare-earth hexaborides even without considering the Kondo mechanism.

\bigskip
\bigskip
\textbf{Acknowledgements}
\newline

M. N. and S.-Y. X. contributed equally to this work.
Work at Princeton University is supported by the US National Science Foundation Grant, NSF-DMR-1006492. M. Z. H. acknowledges visiting-scientist support from Lawrence Berkeley National Laboratory and additional partial support from the A. P. Sloan Foundation and NSF-DMR-0819860. H. L. acknowledges the Singapore National Research Foundation for support under NRF Grant No. NRF-NRFF2013-03. T.-R. C. and H.-T. J. are supported by the National Science Council, Taiwan. H.-T. J. also thanks NCHC, CINC-NTU, and NCTS, Taiwan, for technical support. T. D. at LANL acknowledges support from the U.S. Department of Energy, Office of Basic Energy Sciences, Division of Material Sciences, and the LANL LDRD program.

\clearpage

\end{document}